\documentclass[traditabstract,longauth]{aa} 
\usepackage{graphicx}
\usepackage{txfonts}
\usepackage{natbib}
\usepackage{array}
\usepackage{url}
\begin{document}

\title{LOFAR detections of low-frequency radio recombination lines towards
Cassiopeia~A}

\author{
    Ashish~Asgekar\inst{\ref{astron}}\and 
    J.~B.~R.~Oonk\inst{\ref{astron}} \and 
    S.~Yatawatta\inst{\ref{astron} \and \ref{kapteyn}} \and 
    R.~J.~van Weeren\inst{\ref{leiden} \and \ref{astron} \and \ref{cfa}} \and 
    J.~P.~McKean\inst{\ref{astron}} \and
    G.~White\inst{\ref{stfc} \and \ref{ou}} \and
    N.~Jackson\inst{\ref{jod}} \and 
    J.~Anderson\inst{\ref{mpifr}}\and
    I.~M.~Avruch\inst{\ref{sron} \and\ref{kapteyn} \and\ref{astron} } \and
    F.~Batejat\inst{\ref{oso}}\and
    R.~Beck\inst{\ref{mpifr}}\and
    M.~E.~Bell\inst{\ref{caastro} \and\ref{soton} } \and
    M.~R.~Bell\inst{\ref{mpifa}}\and
    I.~van Bemmel\inst{\ref{astron}}\and
    M.~J.~Bentum\inst{\ref{astron}}\and
    G.~Bernardi\inst{\ref{kapteyn}}\and
    P.~Best\inst{\ref{roe}}\and
    L.~B\^{i}rzan\inst{\ref{leiden}}\and
    A.~Bonafede\inst{\ref{hamburg}}\and
    R.~Braun\inst{\ref{csiro}}\and
    F.~Breitling\inst{\ref{aip}}\and
    R.~H.~van de Brink\inst{\ref{astron}}\and
    J.~Broderick\inst{\ref{soton}}\and
    W.~N.~Brouw\inst{\ref{astron} \and\ref{kapteyn} } \and
    M.~Br\"uggen\inst{\ref{hamburg}}\and
    H.~R.~Butcher\inst{\ref{astron} \and\ref{anu} } \and
    W.~van Cappellen\inst{\ref{astron}}\and
    B.~Ciardi\inst{\ref{mpifa}}\and
    J.~E.~Conway\inst{\ref{oso}}\and
    F.~de Gasperin\inst{\ref{hamburg}}\and
    E.~de Geus\inst{\ref{astron}}\and
    A.~de Jong\inst{\ref{astron}}\and
    M.~de Vos\inst{\ref{astron}}\and
    S.~Duscha\inst{\ref{astron}}\and
    J.~Eisl\"offel\inst{\ref{tls}}\and
    H.~Falcke\inst{\ref{nijmegen} \and\ref{astron} } \and
    R.~A.~Fallows\inst{\ref{astron}}\and
    C.~Ferrari\inst{\ref{nice}}\and
    W.~Frieswijk\inst{\ref{astron}}\and
    M.~A.~Garrett\inst{\ref{astron} \and\ref{leiden} } \and
    J-M.~Grie\ss{}meier\inst{\ref{cnrs} \and\ref{nancay} } \and
    T.~Grit\inst{\ref{astron}}\and
    A.~W.~Gunst\inst{\ref{astron}}\and
    T.~E.~Hassall\inst{\ref{soton} \and\ref{jod} } \and
    G.~Heald\inst{\ref{astron}}\and
    J.~W.~T.~Hessels\inst{\ref{astron} \and\ref{uva} } \and
    M.~Hoeft\inst{\ref{tls}}\and
    M.~Iacobelli\inst{\ref{leiden}}\and
    H.~Intema\inst{\ref{leiden} \and\ref{nrao} } \and
    E.~Juette\inst{\ref{raiub}}\and
    A.~Karastergiou\inst{\ref{ox}}\and
    J.~Kohler\inst{\ref{mpifr}}\and
    V.~I.~Kondratiev\inst{\ref{astron} \and\ref{lebedev} } \and
    M.~Kuniyoshi\inst{\ref{mpifr}}\and
    G.~Kuper\inst{\ref{astron}}\and
    C.~Law\inst{\ref{berkeley} \and\ref{uva} } \and
    J.~van Leeuwen\inst{\ref{astron} \and\ref{uva} } \and
    P.~Maat\inst{\ref{astron}}\and
    G.~Macario\inst{\ref{nice}}\and
    G.~Mann\inst{\ref{aip}}\and
    S.~Markoff\inst{\ref{uva}}\and
    D.~McKay-Bukowski\inst{\ref{stfc} \and\ref{sodankyla} } \and
    M.~Mevius\inst{\ref{astron} \and\ref{kapteyn} } \and
    J.~C.~A.~Miller-Jones\inst{\ref{curtin} \and\ref{uva} } \and
    J.~D.~Mol\inst{\ref{astron}}\and
    R.~Morganti\inst{\ref{astron} \and\ref{kapteyn} } \and
    D.~D.~Mulcahy\inst{\ref{mpifr}}\and
    H.~Munk\inst{\ref{astron}}\and
    M.~J.~Norden\inst{\ref{astron}}\and
    E.~Orru\inst{\ref{astron} \and\ref{nijmegen} } \and
    H.~Paas\inst{\ref{groningen}}\and
    M.~Pandey-Pommier\inst{\ref{lyon}}\and
    V.~N.~Pandey\inst{\ref{astron}}\and
    R.~Pizzo\inst{\ref{astron}}\and
    A.~G.~Polatidis\inst{\ref{astron}}\and
    W.~Reich\inst{\ref{mpifr}}\and
    H.~R\"ottgering\inst{\ref{leiden}}\and
    B.~Scheers\inst{\ref{uva} \and\ref{cwi} } \and
    A.~Schoenmakers\inst{\ref{astron}}\and
    J.~Sluman\inst{\ref{astron}}\and
    O.~Smirnov\inst{\ref{astron} \and\ref{crat} \and\ref{skasa} } \and
    C.~Sobey\inst{\ref{mpifr}}\and
    M.~Steinmetz\inst{\ref{aip}}\and
    M.~Tagger\inst{\ref{cnrs}}\and
    Y.~Tang\inst{\ref{astron}}\and
    C.~Tasse\inst{\ref{meudon}}\and
    R.~Vermeulen\inst{\ref{astron}}\and
    C.~Vocks\inst{\ref{aip}}\and
    R.~A.~M.~J.~Wijers\inst{\ref{uva}}\and
    M.~W.~Wise\inst{\ref{astron} \and\ref{uva} } \and
    O.~Wucknitz\inst{\ref{mpifr} \and\ref{ubonn} } \and
    P.~Zarka\inst{\ref{meudon}}
} 

\institute{
    Netherlands Institute for Radio Astronomy (ASTRON), Postbus 2, 7990 AA Dwingeloo, The Netherlands \label{astron} 
    \and
    Kapteyn Astronomical Institute, PO Box 800, 9700 AV Groningen, The Netherlands \label{kapteyn} 
    \and
    Leiden Observatory, Leiden University, PO Box 9513, 2300 RA Leiden, The Netherlands \label{leiden} 
    \and
    STFC Rutherford Appleton Laboratory,  Harwell Science and Innovation Campus,  Didcot  OX11 0QX, UK \label{stfc} 
    \and
    International Centre for Radio Astronomy Research - Curtin University, GPO Box U1987, Perth, WA 6845, Australia \label{curtin} 
    \and
    University of Hamburg, Gojenbergsweg 112, 21029 Hamburg, Germany \label{hamburg} 
    \and
    Institute for Astronomy, University of Edinburgh, Royal Observatory of Edinburgh, Blackford Hill, Edinburgh EH9 3HJ, UK \label{roe} 
    \and
    Harvard-Smithsonian Center for Astrophysics, 60 Garden Street, Cambridge, MA 02138, USA \label{cfa} 
    \and
    Research School of Astronomy and Astrophysics, Australian National University, Mt Stromlo Obs., via Cotter Road, Weston, A.C.T. 2611, Australia \label{anu} 
    \and
    Department of Astrophysics/IMAPP, Radboud University Nijmegen, P.O. Box 9010, 6500 GL Nijmegen, The Netherlands \label{nijmegen} 
    \and
    Onsala Space Observatory, Dept. of Earth and Space Sciences, Chalmers University of Technology, SE-43992 Onsala, Sweden \label{oso} 
    \and
    Sodankyl\"{a} Geophysical Observatory, University of Oulu, T\"{a}htel\"{a}ntie 62, 99600 Sodankyl\"{a}, Finland \label{sodankyla} 
    \and
    Astronomical Institute 'Anton Pannekoek', University of Amsterdam, Postbus 94249, 1090 GE Amsterdam, The Netherlands \label{uva} 
    \and
    Centre for Radio Astronomy Techniques \& Technologies (RATT), Department of Physics and Elelctronics, Rhodes University, PO Box 94, Grahamstown 6140, South Africa \label{crat} 
    \and
    SRON Netherlands Insitute for Space Research, Sorbonnelaan 2, 3584 CA, Utrecht, The Netherlands \label{sron} 
    \and
    Centre de Recherche Astrophysique de Lyon, Observatoire de Lyon, 9 av Charles Andr\'{e}, 69561 Saint Genis Laval Cedex, France \label{lyon} 
    \and
    SKA South Africa, 3rd Floor, The Park, Park Road, Pinelands, 7405, South Africa \label{skasa} 
    \and
    School of Physics and Astronomy, University of Southampton, Southampton, SO17 1BJ, UK \label{soton} 
    \and
    Centrum Wiskunde \& Informatica, PO Box 94079, 1090 GB,Amsterdam \label{cwi} 
    \and
    Laboratoire Lagrange, UMR7293, Universit\`{e} de Nice Sophia-Antipolis, CNRS, Observatoire de la C\'{o}te d'Azur, 06300 Nice, France \label{nice} 
    \and
    LESIA, UMR CNRS 8109, Observatoire de Paris, 92195   Meudon, France \label{meudon} 
    \and
    Center for Information Technology (CIT), University of Groningen, The Netherlands \label{groningen} 
    \and
    Laboratoire de Physique et Chimie de l' Environnement et de l' Espace, LPC2E UMR 7328 CNRS, 45071 Orl\'{e}ans Cedex 02, France \label{cnrs} 
    \and
    Radio Astronomy Lab, UC Berkeley, CA, USA \label{berkeley} 
    \and
    Jodrell Bank Center for Astrophysics, School of Physics and Astronomy, The University of Manchester, Manchester M13 9PL,UK \label{jod} 
    \and
    Argelander-Institut f\"{u}r Astronomie, University of Bonn, Auf dem H\"{u}gel 71, 53121, Bonn, Germany \label{ubonn} 
    \and
    Astro Space Center of the Lebedev Physical Institute, Profsoyuznaya str. 84/32, Moscow 117997, Russia \label{lebedev} 
    \and
    Th\"{u}ringer Landessternwarte, Sternwarte 5, D-07778 Tautenburg, Germany \label{tls} 
    \and
    Max Planck Institute for Astrophysics, Karl Schwarzschild Str. 1, 85741 Garching, Germany \label{mpifa} 
    \and
    Leibniz-Institut f\"{u}r Astrophysik Potsdam (AIP), An der Sternwarte 16, 14482 Potsdam, Germany \label{aip} 
    \and
    National Radio Astronomy Observatory, 520 Edgemont Road, Charlottesville, VA 22903-2475, USA \label{nrao} 
    \and
    Max-Planck-Institut f\"{u}r Radioastronomie, Auf dem H\"ugel 69, 53121 Bonn, Germany \label{mpifr} 
    \and
    Astrophysics, University of Oxford, Denys Wilkinson Building, Keble Road, Oxford OX1 3RH \label{ox} 
    \and
    Astronomisches Institut der Ruhr-Universit\"{a}t Bochum, Universitaetsstrasse 150, 44780 Bochum, Germany \label{raiub} 
    \and
    ARC Centre of Excellence for All-sky astrophysics (CAASTRO), Sydney Institute of Astronomy, University of Sydney Australia \label{caastro} 
    \and
    Station de Radioastronomie de Nan\c{c}ay, Observatoire de Paris, CNRS/INSU, 18330 Nan\c{c}ay, France \label{nancay} 
    \and
    CSIRO Australia Telescope National Facility, PO Box 76, Epping NSW 1710, Australia \label{csiro} 
    \and
     Department of Physics \& Astronomy, The Open University, UK \label{ou} 
}

\date{Received December 24, 2012; accepted February 11, 2013}

\titlerunning{RRLs towards Cas A using LOFAR}
\authorrunning{Asgekar et al.}

\abstract{

Cassiopeia A was observed using the Low-Band Antennas of the LOw Frequency ARray (LOFAR)
with high spectral resolution. This allowed a search for
radio recombination lines (RRLs) along the line-of-sight to this source. Five carbon $\alpha$
RRLs were detected in absorption between 40 and 50~MHz with a signal-to-noise ratio
of $>5$ from two independent LOFAR datasets.  
The derived line velocities ($v_{\rm LSR} \sim-50$~km\,s$^{-1}$)
and integrated optical depths ($\sim 13$~s$^{-1}$) of the RRLs
in our spectra, extracted over the whole supernova remnant, are
consistent within each LOFAR dataset and with those previously reported. For the first time, we are able
to extract spectra against the brightest hotspot of the remnant at frequencies
below 330~MHz. These spectra show significantly higher (15--80 percent) integrated optical
depths, indicating that there is small-scale angular structure on the order of $\sim$~1~pc
in the absorbing gas distribution over the face of the
remnant. We also place an upper limit
of 3\,$\times$\,10$^{-4}$ on the peak optical depths of hydrogen and helium RRLs.
These results demonstrate that LOFAR has the desired spectral stability
and sensitivity to study faint recombination lines in the decameter band. 
}

\keywords{ISM: clouds -- radio lines : ISM -- ISM: individual objects: Cassiopeia~A}
\maketitle


\section{Introduction}

Radio recombination lines (RRLs) arise from ions recombining with electrons 
in diffuse, partially ionised gas. At high principal quantum numbers (i.e. $n\ga$~200 at 
frequencies $\la$~1~GHz), RRLs are an important probe of the temperature and density of 
the cool ($T\la$~100~K) interstellar medium \citep[e.g.,][]{Payne_etal:1989, Konovalenko:1984}.
RRL measurements have distinct advantages over
other tracers of the ionised gas; the extinction caused by dust and the contamination
from other sources are negligible.
However, there have been only a limited number of surveys that have studied RRLs
from carbon at low radio frequencies \citep[$\la$~330~MHz;
e.g.][]{Erickson95,Kantharia_Anantha:2001JApA22,Roshi_etal:2002AA391}.
Also, the low angular resolution of these early surveys, from several
degrees to around 6~arcmin, did not allow the location of the line-emitting
gas to be determined, and also limited their capability to detect faint
RRL sources due to beam dilution effects.

The LOw Frequency ARray (LOFAR; \citealt{stappers12,haarlem13}) is a new-generation 
radio telescope, which given its unique design, can be used to make 
sensitive surveys for RRLs with high spectral and spatial resolution.
LOFAR comprises thousands of antennas arranged in (currently) 41~stations
spread across the Netherlands and other European countries. By design,
LOFAR provides excellent baseline coverage, particularly in the inner 3~km
core-region, where over half of the collecting area is contained. The low
band antennas (LBAs) of LOFAR, although optimised for frequencies above 30~MHz,
cover the frequency range from 10 to 90~MHz. The high band antennas (HBAs)
cover the frequency range from 110 to 250~MHz. This large observable frequency
window makes LOFAR an ideal instrument for carrying out studies of the ISM
with low-frequency RRLs.

In this Letter, we present the first spectral line measurements made using the
LBA system of LOFAR. We target carbon RRLs in absorption towards the supernova
remnant Cassiopeia~A (Cas A), which is one of the brightest continuum sources
in the sky and has a sightline that is already known to display strong  carbon
RRLs in low frequency spectra \citep[see][and  
references therein]{Stepkin_etal:2007}. In Sect. \ref{sec:data}, we present our
LOFAR commissioning observations and data analysis steps. Our carbon RRL detections
are presented in Sect. \ref{results}, and we discuss these detections and present
our conclusions in Sect. \ref{sec:disc}.


\section{Observations \& Data Reduction}
\label{sec:data}

\begin{table*}
\caption{Summary of the LOFAR data sets used for our analysis: raw visibilities
    from Data-1 were processed twice to create two separate intermediate data sets for 
    i)  continuum imaging and 
    ii) the RRL detections (see Section~\ref{sec:data}).
    The velocity resolutions were computed at 45~MHz.}
\begin{tabular}{llllllll}
\hline
Data set & LOFAR id & Start date &  Time & Duration & \multicolumn{3}{c}{Processed resolution} \\
&&&(UT)&(h)&Int. time (s) & Channel width (kHz) & Channel width (km\,s$^{-1}$)\\
\hline
Data-1 (continuum)	& L25937 & 2011 May 4	& 23:00:01.5	& 18	& 5 &  48.768 & 325\\
Data-1 (RRL)		& L25937	& 2011 May 4	& 23:00:01.5	& 18	& 5 & 0.762 & 5.1\\ 
Data-2 (RRL)		& L31848 & 2011 Oct 15	& 12:00:02.5	& 15	& 5 & 3.1 &	21 \\\hline
\end{tabular}
\label{tab:data}
\end{table*}

The data sets for our RRL study are taken from two LOFAR commissioning observations of Cas A in the 30 to 90~MHz (LBA) band. The first data set (Data-1) was taken on 2011 May 4 using the 16 core stations and 7 remote stations that were constructed at the time, giving baselines up to about 24 km. These data were taken with 1-s visibility integrations and 244 separate frequency subbands, each with a bandwidth of 195 kHz and 256 spectral channels. The second data set (Data-2) was observed on 2011 October 15 using 24 core stations and 8 remote stations, and had a maximum baseline length of about 83 km. Data-2 was taken in a multi-beam mode that is used for standard continuum observations, with one beam on Cas A and a second beam on Cygnus A. This resulted in only half of the available bandwidth being used for the observation of Cas A. Also, a coarser spectral resolution of 64 spectral channels within each of the 122 subbands of 195 kHz bandwidth was used. The visibility integration time was again 1 s. After the first step of
interference removal using the AOFlagger routine \citep[][]{Offringa_etal:2010MNRAS},
the data were averaged in time and/or frequency as required.

In total, we produced 3 data sets for our work. First, to make a continuum image of Cas A, we produced an intermediate data set from Data-1 that had a coarse spectral resolution of $\sim$~50~kHz~channel$^{-1}$ in order to speed up the calibration process later. Next, we created two RRL data sets. For Data-1, we included
only those baselines between the separate core-stations ($<$~2.6~km) and retained the full spectral resolution 
of 762~Hz~channel$^{-1}$. Data-2 was processed to retain
a spectral resolution of 3.1~kHz~channel$^{-1}$ and to also include longer baselines
($<$~24~km) so that spectra could be extracted from individual regions across the supernova remnant. For these two RRL data sets, we decided to concentrate our search to those  frequencies between 40 to 50 MHz. This was because the RRL absorption was expected to be stronger at the lower end of the LBA spectrum due to
the steep broadband spectral energy distribution of Cas A. The lower frequency end of the search window was chosen because the level of terrestrial interference
increases below 40~MHz \citep[][]{Offringa_etal:2012}. The resulting time and frequency resolution for each data set after these preprocessing steps are summarized in Table~\ref{tab:data}.

\begin{figure}
\begin{center}
\setlength{\unitlength}{1cm}
\begin{picture}(6,7.9)
\put(9.3,-0.6){\includegraphics{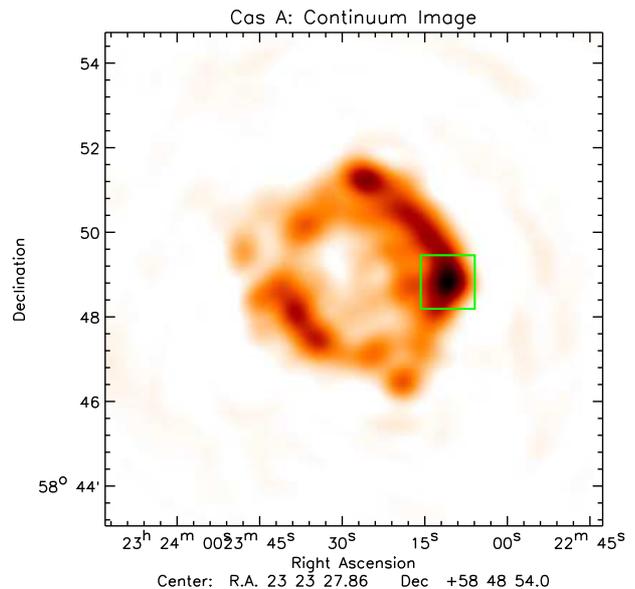}}
\end{picture}
\caption{LOFAR image of Cas A at 52 MHz (single subband of 195 kHz bandwidth). The image
was made using Briggs weighting (robust  = 0) and was restored using a beam size
of $40$~arcsec. The hotspot region used to extract spectra from Data-2 is marked with the
green square.} 
\label{fig:casa_lba_cont}
\end{center}
\end{figure}

We initially calibrated our continuum data using the
BlackBoard Self-calibration package \citep[{\sc bbs};][]{Pandey_etal:2009},
with a relatively simple initial model, containing about a dozen point sources
and one shapelet model. We then used the {\sc sagecal} package \citep[][]{Kazemi_etal:2011MNRAS}
to solve for the direction-dependent amplitude and phase variations in the data
and to update the shapelet model for Cas A. To mitigate artefacts due to Cygnus~A,
present about 30 degrees away, we simultaneously solved for direction-dependent gains
towards Cas A and Cygnus~A, and subtracted the latter in the visibility
data using a shapelet model. We then imaged the source using the {\sc casa}
imager (\url{http://casa.nrao.edu/}) and constructed a
composite sky model from the image by fitting for point and shapelet
sources \citep[see e.g.,][]{Yatawatta:2011}. We 
iterated over this scheme ({\sc bbs} calibration $\rightarrow$ imaging
$\rightarrow$ skymodel $\rightarrow$ calibration) to obtain high-resolution
continuum images and improved source models. 
In Fig.~\ref{fig:casa_lba_cont}, we present a 40~arcsec resolution
continuum image of the supernova remnant at 52 MHz (single subband). Given the 
excellent spatial resolution of LOFAR, a number of complex features are clearly resolved
\citep[c.f.,][]{Kantharia_etal:1998,Lane_etal:2005}. Further details 
of the calibration strategy and a full discussion of the
continuum imaging of Cas A, and the wider surrounding field, will be presented in a
companion paper (Yatawatta et al., in prep).

To detect RRLs along the line-of-sight to Cas A, we calibrated the RRL visibility data
sets (see Table~\ref{tab:data}) with {\sc bbs} using our final sky model
that was obtained from our analysis of the continuum data. Multi-channel maps
were constructed using the {\sc clean} algorithm within {\sc casa},
and image cubes from the individual subbands were analysed separately. Spectra were
then extracted over the whole remnant and over the brightest hotspot (see Fig. \ref{fig:casa_lba_cont}). The resultant spectra displayed
a few artefacts, such as breaks at the subband boundaries and bad channels
on the edge of each subband; typically around 5 per cent of the channels per subband.
These bad channels were removed during further analysis. Doppler correction
terms are currently not
incorporated in the LOFAR imaging pipeline, hence the spectra were Doppler corrected
after extraction. This leads to an uncertainty of $\la$~1~km\,s$^{-1}$ in the
determination of the central velocities and line widths. 
This error is negligible compared to the velocity resolution of our data.


\section{Results}
\label{results}

\begin{figure}
\begin{center}
\setlength{\unitlength}{1cm}
\begin{picture}(6.0,6.1)
\put(7.35,-0.8){\includegraphics{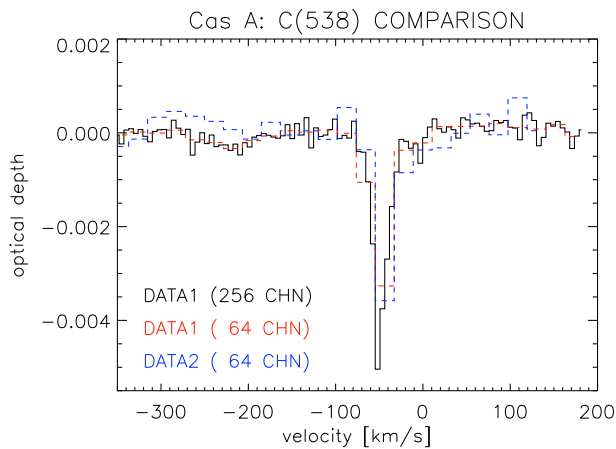}}
\end{picture}
\caption{A representative detection of a carbon $\alpha$ RRL towards Cas A with LOFAR. We show spectra of the RRL C538$\alpha$ that was measured using Data-1 (black solid) and Data-2 (blue dashed). For comparison, we have also smoothed the Data-1 spectrum to the same spectral resolution as Data-2 (red dashed). These spectra demonstrate the spectral stability of the LOFAR system for making spectral line measurements.} 
\label{fig:casa_rrl1}
\end{center}
\end{figure}

\begin{figure}
\begin{center}
\setlength{\unitlength}{1cm}
\begin{picture}(6.0,5.9)
\put(8.25,-0.9){\includegraphics{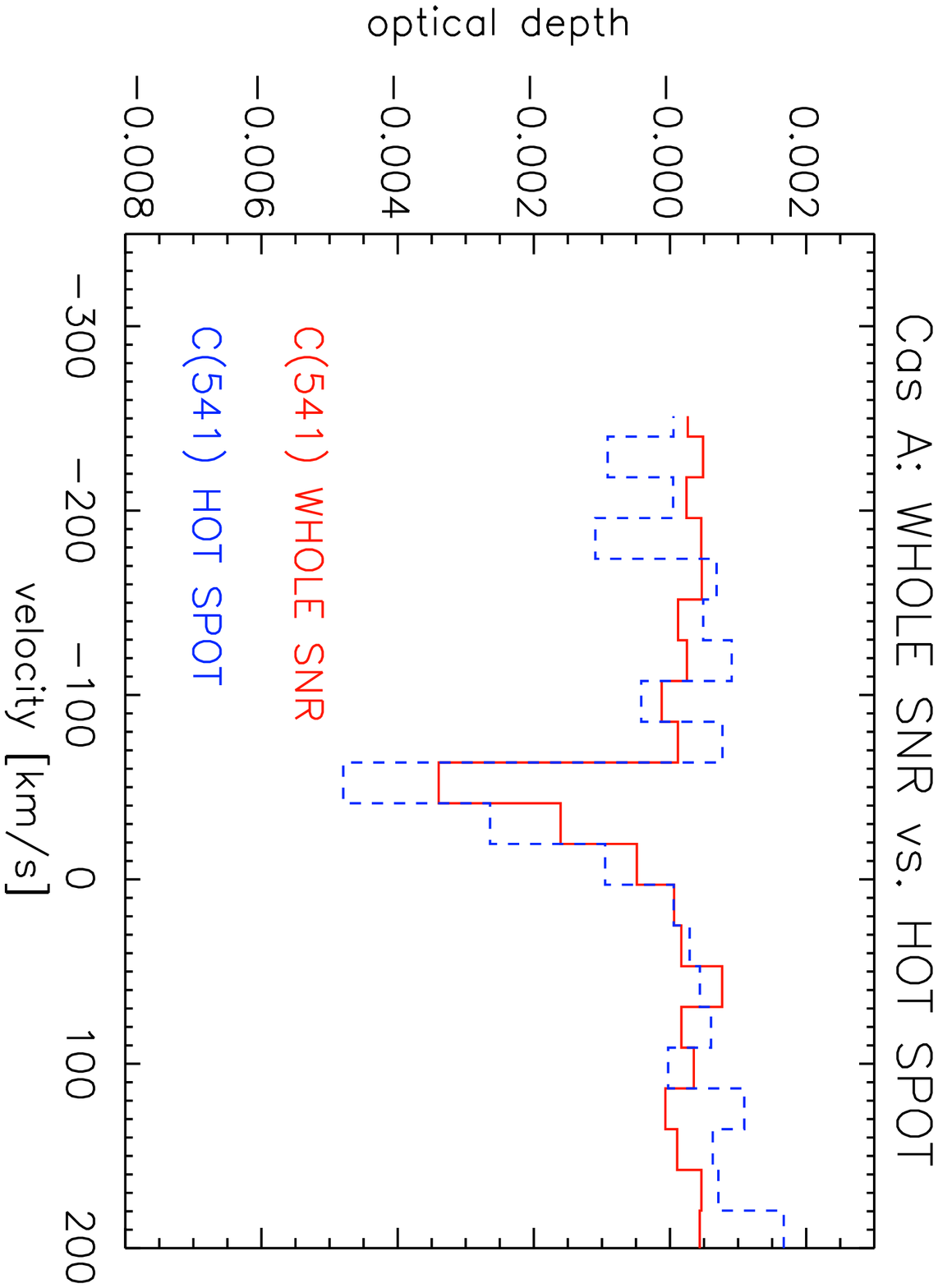}}
\end{picture}
\caption{The spectrum of the RRL C541$\alpha$ measured over the whole remnant (red) and only over the hotspot region (blue) using Data-2. These data demonstrate that the integrated optical depth over the hotspot region is higher than across the remnant, showing that there is likely structure in the absorbing gas on scales of $\sim$~1~pc.}
\label{fig:casa_rrl2}
\end{center}
\end{figure}

The subset of Data-1 that was used for the RRL analysis only contained the short baselines between the
core-stations ($\la$~2.6~km), for which Cas A is practically unresolved. Therefore, spectra were
obtained over the whole remnant with a spectral resolution of 762~Hz~channel$^{-1}$, which is
equivalent to a velocity resolution of $\sim\,5$~km\,s$^{-1}$ at 45~MHz.
We report the detection (at $>$\,16$\sigma$ level) of four carbon $\alpha$ RRLs (C518$\alpha$,
C538$\alpha$, C541$\alpha$ and C543$\alpha$) towards Cas A in Data-1 between 40 to 50 MHz. In
Fig.~\ref{fig:casa_rrl1}, we present a representative spectrum of a RRL detection, 
with respect to the local standard of rest (LSR). For each detected RRL, we measured the line
widths and integrated optical depths ($\int \tau_\nu\,d\nu$) by fitting Gaussian components to the spectra,
the results of which are presented in Table~\ref{tab:rrls}. Only single Gaussian-line profiles were used because
our spectra do not currently have a sufficient signal-to-noise ratio to reliably measure multiple
components in the line structure. All of the detected RRLs have measured velocities at
$v_{\rm LSR} \sim -48\pm 1$~km\,s$^{-1}$, consistent with the velocity
of the gas in the Perseus arm, the likely absorber along the line-of-sight to Cas A
\citep[][]{Konovalenko:1984, Ershov_etal:1984, Ershov_etal:1987, Payne_etal:1989, Payne_etal:1994}.
These RRL detections from Data-1 represent the first spectral-line measurements made with LOFAR.

The main goal of our study is to investigate whether there is any structure in the absorbing gas
by detecting a variation in the integrated optical depths of the RRLs over the extent of the remnant.
For this, LOFAR's unparalleled spatial resolution at frequencies below $100$~MHz is required. We extracted
spectra using Data-2 in two ways; one towards the compact hotspot component and the other over the
entire area of the remnant. We detected five carbon $\alpha$ RRLs (at $>$~5$\sigma$ level) in both
our integrated spectra and the spectra extracted over the brightest hotspot component in the remnant.
As these lines were only marginally resolved, we estimated their integrated optical depths by summing
the measured optical depth over two channels. The results of these measurements are also given in
Table~\ref{tab:rrls}. In Fig.~\ref{fig:casa_rrl2}, we show a representative detection of a
carbon $\alpha$ RRL (C541$\alpha$) against the whole remnant and the hotspot.
Earlier studies \citep[see e.g.,][]{Payne_etal:1989} reported two components
at $v_{\rm LSR} \sim-50$ and $\sim-40$~km\,s$^{-1}$, where the former dominates by a
factor~$2-3$. Due to insufficient signal-to-noise ratio we can not currently separate
these components, but the observed line-profile asymmetry in our spectra (see 
Fig.s~\ref{fig:casa_rrl1}~and~\ref{fig:casa_rrl2}) is consistent with two components.

We also carried out a search for hydrogen and helium RRLs in the individual
subbands of our data sets (without any folding in frequency), but found no significant
detection. The spectral noise per channel in our most sensitive data set
(Data-1) is $\sim$~10$^{-4}$, so we derive an upper limit (3$\sigma$) on the peak optical
depth of hydrogen and helium RRLs of 3\,$\times$\,10$^{-4}$, or an equivalent integrated
optical depth of 0.9~s$^{-1}$. 
Following \citet[][]{Shaver:1976}, we derive an upper limit 
of $4\times 10^{-18}\,\mbox{s}^{-1}$ for the interstellar ionisation rate of hydrogen.
Our limit is significantly lower than the previously reported values despite the
uncertainties in determining various parameters~\citep[see e.g.,][\S 3.3.6]{Payne_etal:1994,
Gordon_Sorochenko:2009}, indicating that cosmic rays can not ionise the intercloud
gas to the observed degree.


\section{Discussion \& Conclusions}
\label{sec:disc}
\begin{figure}
\begin{center}
\setlength{\unitlength}{1cm}
\begin{picture}(6,6.4)
\put(8.0,-0.8){\includegraphics{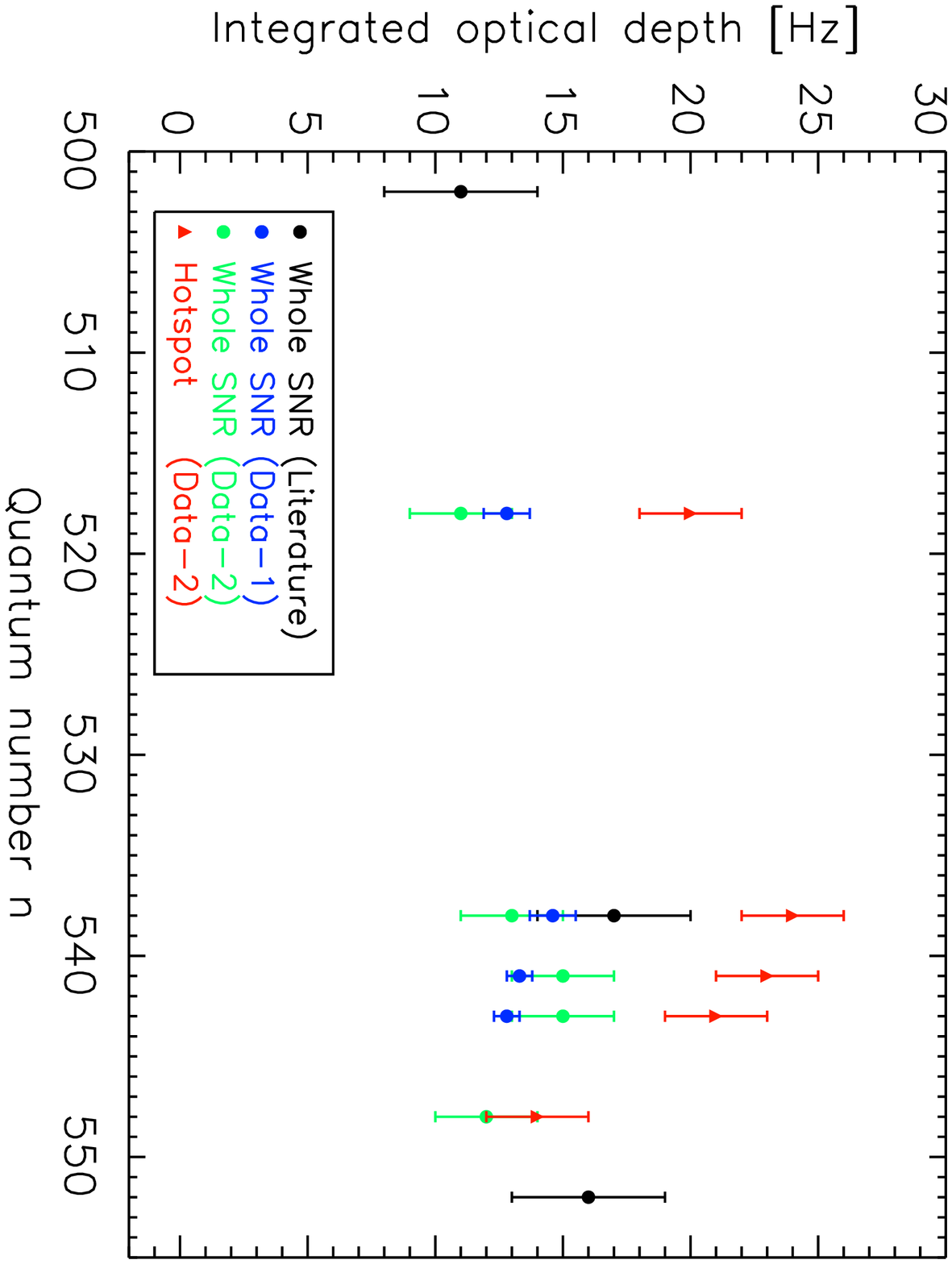}}
\end{picture}
\caption{The integrated optical depths as a function of the principal quantum number for
both data sets (Data-1 and Data-2) and for those previously
reported in the literature \citep{Ershov_etal:1984,Ershov_etal:1987,Payne_etal:1989,Payne_etal:1994}.
We find good agreement between the measurements made over the whole remnant from the two
LOFAR data sets and with those in the literature. We also find that the integrated optical
depths of the RRLs are higher over the bright hotspot component with respect to the measurements
over the whole remnant, implying that there is likely small-scale structure in the absorbing gas distribution.} 
\label{fig:optdepths}
\end{center}
\end{figure}

In total, we have detected five carbon $\alpha$ RRLs towards Cas A from our two 
LOFAR commissioning data sets. In Fig.~\ref{fig:optdepths}, we plot the integrated
optical depths of these RRLs as measured from each 
LOFAR data set, and also the
integrated
optical depths for those RRLs that were previously measured towards Cas A with
other telescopes
~\citep[][]{Ershov_etal:1984,Ershov_etal:1987,Payne_etal:1989,Payne_etal:1994},
as a function of the principal quantum number. We see that the integrated optical depths,
as measured over the whole remnant, are consistent between Data-1 and Data-2 for the
four RRLs that are common to both data sets. This demonstrates the spectral stability
of our LOFAR commissioning data sets. We also find consistency for the RRL C538$\alpha$,
where we have both LOFAR and previously reported measurements in the literature. Although
we currently do not have measurements for the other two previously detected RRLs 
(C502$\alpha$ and C552$\alpha$), the reported integrated optical depths of these lines
are generally consistent with the RRLs that we have detected in the LOFAR data sets.

Multiple carbon RRLs are usually modelled together in order to derive the physical
conditions of the line-absorbing gas, since the electron levels are in non-local thermal
equilibrium (non-LTE) conditions. Crucial inputs to these models include accurate line
width measurements from high-spectral resolution data. Kantharia et al. (1998) show that
previous low-frequency measurements of carbon $\alpha$ RRLs can be modelled as originating
in cold ($T_{e}\sim75$~K), low density ($n_{e}\sim0.02$~cm$^{-3}$) clouds that are located
approximately $115$~pc away from Cas~A. Higher density models ($T_{e}\sim75$~K, $n_{e}\sim0.1$~cm$^{-3}$)
are also able to fit the data, but these are less appealing due to their thermal pressures
being very high. Although our current RRL detections are consistent with these models,
given the agreement with the previous measurements, future observations of RRLs over the
whole LOFAR observing window, and hence over a large and densely sampled range of principal
quantum number, will differentiate between these competing models. The results from such a
detailed analysis using new observations will be reported in a forthcoming paper (Oonk et al., in prep).

\begin{table}
\tabcolsep=0.12cm
\caption{The derived line parameters for the RRLs measured from Data-1 and Data-2. For the 
case of Data-1, the spectral resolution was sufficient to resolve the lines and the fitted
Gaussian full width at half maximum (FWHM) is reported.  For Data-2, the RRLs were not fully
resolved and so the integrated optical depths were obtained over two spectral channel widths.
The integrated optical depths have been calculated over the full remnant (SNR) for datasets
Data-1 and Data-2, and around the region of the bright hotspot for Data-2 (see Fig.~\ref{fig:casa_lba_cont}).}
\begin{tabular}{lllllll}
\hline 
Line	     & Rest 				& Center  & FWHM 			&\multicolumn{3}{c}{Integrated optical depth}\\
          & frequency& velocity & (km\,s$^{-1}$)	& \multicolumn{3}{c}{(s$^{-1}$)} \\
          & (MHz)				& (km\,s$^{-1}$) & Data-1			& Data-1	& \multicolumn{2}{c}{Data-2} \\
          & 					    & Data-1 & 				& SNR	& SNR	&Hotspot \\
\hline
C548$\alpha$	& 39.87	&   $\;\;$--     &		$\;\;$--					&	$\;\;$--			& 12\,$\pm$\,2	& 14\,$\pm$\,2 \\
C543$\alpha$	& 40.98	&$-48\pm1$&			 19\,$\pm$\,2	& 12.8\,$\pm$\,0.5	& 15\,$\pm$\,2	& 21\,$\pm$\,2 \\
C541$\alpha$	& 41.44	&$-48\pm1$&			 19\,$\pm$\,2	& 13.3\,$\pm$\,0.5	& 15\,$\pm$\,2	& 23\,$\pm$\,2 \\
C538$\alpha$	& 42.13	&$-48\pm1$&			 20\,$\pm$\,3	& 14.6\,$\pm$\,0.9	& 13\,$\pm$\,2	& 24\,$\pm$\,2 \\
C518$\alpha$	& 47.20	&$-48\pm1$&			 18\,$\pm$\,2	& 12.8\,$\pm$\,0.9	& 11\,$\pm$\,2	& 20\,$\pm$\,2 \\
\hline
\end{tabular}
\label{tab:rrls}
\end{table}

We also find from Fig.~\ref{fig:optdepths} that the integrated optical depths towards the
hotspot are larger by up to almost a factor of 2, when compared to the integrated optical
depths measured over the whole remnant. This shows that there is small-scale angular
structures and/or the geometry of the RRL-absorbing gas is such that there is varying
levels of absorption over the face of the remnant \citep[see, e.g.][]{Kantharia_etal:1998}.
This corresponds to structure in the RRL-absorbing gas on the order
of $\sim$~1~pc. To model this structure, we plan to make new LOFAR observations
with a higher spectral and spatial resolution in the future.

Our results from the commissioning data sets demonstrate that LOFAR can be used to carry
out spectral line observations at low radio frequencies. For the case of RRL studies, LOFAR
will be an important tool in the future for two main reasons. First, given the up to
arcsecond-scale resolution that can be obtained with LOFAR, beam dilution effects will not 
affect our detection sensitivity. Second, due to the large observable frequency range of
LOFAR, from 10 to 250 MHz, hundreds of low-frequency RRLs will be detectable for the first
time. Furthermore, due to the large bandwidth that can be observed during each observation,
many of these RRLs will be detected in a single observation, making wide-field surveys for
RRLs possible for the first time. The results reported here form part of the LOFAR Surveys
key science project. In the near future, we plan to observe various bright sources (Galactic
and extragalactic) with compact morphologies in the LOFAR LBA and HBA bands. Later studies
will focus on RRLs towards regions of diffuse emission in the Galactic plane and fields with
complex emission features.

\begin{acknowledgements}
LOFAR, designed and constructed by ASTRON, has facilities in several countries,
that are owned by various parties (each with their own funding sources), and 
that are collectively operated by the International LOFAR Telescope (ILT)
foundation under a joint scientific policy. AA~would like to thank M.~I.~R. Alves
for discussions and comments on an early version of this manuscript. AA~acknowledges
support from IUCAA, Pune,
during the preparation of this manuscript. RJvW acknowledges support from NASA 
through the Einstein
Postdoctoral grant number PF2-130104 awarded by the Chandra X-ray Center, which is
operated by the Smithsonian Astrophysical Observatory for NASA under
contract NAS8-03060. CF acknowledges financial support by the ``Agence Nationale
de la Recherche" through grant ANR-09-JCJC-0001-01. 
\end{acknowledgements}


\end{document}